\documentclass[preprint]{revtex4-1}
\usepackage{amsmath}
\usepackage{amsfonts}
\usepackage{amssymb}
\usepackage{graphicx}

\begin{document}

\title{Drawing Particle Symbols and Feynman Diagrams with Office}
\author{Orhan \c{C}ak\i r\\Ankara University, Faculty of Science, Department of Physics,
\\06100, Tandogan, Ankara, Turkey.}

\begin{abstract}
Particle symbols and Feynman diagrams often used in particle physics
can be generated by a word document processor. Standard office
packages have their built in symbols such as solid-line,
dashed-line, arrows, etc. which are also used in particle physics.
Three symbols corresponding to photon, gluon and fermion line are
not present in the standard office packages. By importing these
symbols and using the built in symbols, tree-level Feynman diagrams
can be easily generated in an office program.
\end{abstract}
\maketitle

\section{Particle Symbols}

This is a simple guide to generate a Feynman diagram [1] in Openoffice.org [2]
or other office packages [3,4]. Standard office packages have their own
symbols and more can be generated easily by using the drawing toolbar menu.
Furthermore, selecting more than one symbols or drawings can be grouped into a figure.

Some of the particle symbols used in particle physics to draw Feynman diagrams
are not present standard drawing toolbar menu. We need to be supply these
symbols from an external input. Most used symbols in quantum electrodynamics
(QED) and quantum chromodynamics (QCD) are photon and gluon, respectively.
These symbols are shown in Table I.

\bigskip

Table I. Particle symbols not present in standard office packages
and a fermion line composed of an arrow and a line.
\begin{table}[h]
\begin{tabular}[c]{lll}
\hline\hline
Photon (or massive vector boson) &  &
{\includegraphics[ scale=0.6 ] {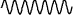} }
\\ \hline
Gluon &  & {\includegraphics[ scale=0.6 ] {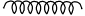} }
\\ \hline
Fermion &  & {\includegraphics[ scale=0.6 ] {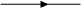}}
\\ \hline\hline
\end{tabular}
\end{table}

\section{Feynman Diagram}

A Feynman diagram corresponding to a process can be formed by
copying and/or rotating the above symbols according to the related
physics model. As an example, a diagram for the process $\gamma
q\rightarrow gq$ is shown in Fig. 1. A different size of these
symbols can be obtained by clicking on the object and pulling out
from the corner.

Orientation of these symbols can be changed by rotating it around
their centers. After a careful drawing and linearization of the
diagram, a publication quality Feynman diagram can be obtained. The
diagram can be converted to other formats such as postscript (PS) or
printer file (PRN) using a postscript printer driver and the
\textquotedblleft print to file\textquotedblright\ facility of the
office package. Furthermore using the program Ghostview [5], the
file with an extension PS or PRN can be converted to an encapsulated
postscript (EPS) or a PNG figure file providing the original file
must be a single page document.

\section{Conclusion}

This simple guide for Feynman diagram drawing in office packages can
be used and distributed without changing its original form. For a
colourful example, see some particle symbols and Feynman digrams in
the APPENDIX A. Upon usage of these symbols, it should be referred
to the preprint number arXiv:physics/0411006v3. The source file can be
obtained from http://80.251.40.59/science.ankara.edu.tr/ocakir/psfdo\_tr.html.

$$
$$

\begin{figure}[h]
\begin{center}
\includegraphics[scale=0.8] {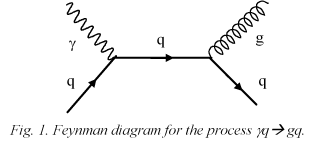}
\end{center}
\end{figure}

The author of the file will not be responsible for the inconveniences from the
wrong use of the program PSFDO.

$$
$$

{\bf References}

[1] Theory of Feynman diagrams, http://www2.slac.stanford.edu/vvc/theory/feynman.html

[2] Openoffice.org Project, it is free to use and distribute, http://www.openoffice.org/

[3] Free office packages, http://www.freebyte.com/office/\#suites

[4] MS Office, http://www.msoffice.com

[5] Ghostview is made available with the Aladdin Free Public Licence,

http://www.ghostgum.com.au/

\newpage

{\bf APPENDIX A}
$$
$$

PSFDO (Particle Symbols and Feynman Diagram with Office programs) FILE V1.0

Particle symbols not present in standard office packages are created
by ORHAN CAKIR. These are shown in Table A-1.

Table A-1. Particle symbols not present in standard office
packages.You can copy, move, rescale all the symbols and diagrams to
make a colourful Feynman diagram.Particle symbols can be grouped
into diagrams shown in Figure A-1 corresponding to a physical
process.

\begin{table}[h]
\begin{tabular}[c]{lll}
\hline\hline Photon (or massive vector boson) &  &
{\includegraphics[ scale=0.6 ] {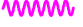} }
\\ \hline
Gluon &  & {\includegraphics[ scale=0.6 ] {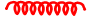} }
\\ \hline
Fermion &  & {\includegraphics[ scale=0.6 ] {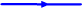} }
\\ \hline\hline
\end{tabular}
\end{table}

$$
$$

\begin{figure}[h]
\begin{center}
\includegraphics[ scale=0.8 ] {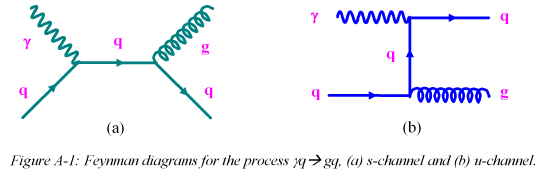}
\end{center}
\end{figure}

{\it Note}: 

Note that the author of the file will not be responsible for the
inconveniences from the wrong use of the program PSFDO.

\end{document}